\documentstyle{article}
\begin{document}
\Large
\title{MINIMAL CLOSED SET OF OBSERVABLES IN THE THEORY \\OF COSMOLOGICAL
PERTURBATIONS}
\author{M.Novello, J.M.Salim, M.C.Motta da Silva, S.E.Jor\'{a}s, R.Klippert \\
Centro Brasileiro de Pesquisas F\'{\i}sicas --- Rio de Janeiro}
\date{}
\maketitle

\begin{abstract}
The theory of perturbation of Friedman-Robertson-Walker (FRW) cosmology is
analysed exclusively in terms of observable quantities. Although this
can be a very complete and general procedure we limit our presentation here to
the case of irrotational perturbations for simplicity. We show that
the electric part of Weyl conformal tensor {\bf $E$} and the shear
{\bf $\Sigma$} constitute the
two basic perturbed variables in terms of which all remaining observable
quantities can be described. Einstein\rq s equations of General Relativity
reduce to a closed set of dynamical system for {\bf $E$} and {\bf $\Sigma$}.
The basis for a gauge-invariant Hamiltonian treatment of the
Perturbation Theory in the FRW background is then set up.
\end{abstract}

\newpage
\large
\section{INTRODUCTION}
\protect\label{introdusao}

\subsection{INTRODUCTORY REMARKS}
\protect\label{remarks}

It has been a common practice (since Lifshitz\rq s original paper
\cite{Lifshitz}) to start
the examination of the Perturbation Theory of Einstein\rq s General
Relativity by considering variations of
non observable quantitites such as $\delta g_{\mu\nu}$. The main
drawback of this procedure is that it mixes true
perturbations and arbitrary (infinitesimal) coordinate transformations.
We are then faced with an extra task: the separation of true
perturbation terms from a mere coordinate transformation. This is the so
called gauge problem of the perturbation theory. A solution for this
difficulty was found by many authors (cf. \cite{Bardeen}, \cite{Jones},
\cite{Hawking}, \cite{Olson}, \cite{Vishniac}, \cite{BrandenbergerKhan},
\cite{MukhanovBrand}) by looking for gauge-independent
combinations which are written in terms of the metric tensor and its
derivatives.

The next step would then be to provide from Einstein\rq s equations,
that deal with $\delta g_{\mu\nu}$,
the dynamics of these gauge-independent variables which
would then be used to describe physically relevant quantities.

Here we will follow a simpler (and more direct) path, inverting this
procedure. That is, we will choose from the beginning, as the basis of our
analysis, the gauge-invariant, physically observable quantities
\footnote{Actually, the gauge problem does not even appear into
our scheme and it can be completely ignored as far as basic physical
quantities of the perturbation theory of FRW geometry are concerned.
That is, there exists a dynamical system that can be analysed without
references to this problem. However, to make contact with the standard
procedure that deals with $\delta g_{\mu\nu}$, we will present the evolution
of these associated gauge-dependent terms later on.}.
The dynamics for these fundamental quantities will then be analysed and
any remaining gauge-dependent objects which we usually deal with will be
obtained from this fundamental set.

There are basically two fundamental
approaches by which the perturbation theory can be elaborated:
one of them makes use of
the standard Einstein\rq s equations \cite{Lifshitz} and the other is based
on the equivalent quasi-Maxwellian description \cite{Hawking}
\cite{Jordan et al} \cite{Novello e Salim}.
In the case of the spatially homogeneous and isotropic FRW
cosmological model the vanishing of Weyl conformal tensor suggests that
the second approach is more attractive. In this case the variation of Weyl
conformal tensor $\delta W_{\alpha\beta\mu\nu}$ is the
basic quantity to be considered, once there is certainly no doubt that
$\delta W_{\alpha\beta\mu\nu}$ is a true perturbation, which can not be
achieved by a coordinate transformation. This solves {\it ab initio} the
gauge problem that was pointed out in the first approach.

The crucial point of distinction between these approaches is that the
dynamics of the observable
quantities, as we shall see, does not require the knowledge of all
components of $\delta g_{\mu\nu}$.

{}From a technical standpoint, instead of considering
tensorial quantities, one should restrain oneself to scalar ones. There are
two ways to implement this:
\begin{itemize}
 \item{Expand the relevant quantities in terms of a complete basis of
functions (e.g. the spherical harmonics basis).}
 \item{Analyse the invariant geometric quantities one can construct from
$g_{\mu\nu}$ and its derivatives in the Riemannian
background structure, that is, examine the 14 Debever invariants.}
\end{itemize}

In any of these ways we shall see that the net result is that there is a set
of perturbed quantities which can be
divided into \mbox{\lq \lq good\rq \rq} quantities (i.e., the ones whose
unperturbed counterparts have zero value in the background and, consequently,
Stewart\rq s lemma \cite{Stewart} guarantees that the associated perturbed
quantity is really a gauge-independent one) and
\mbox{\lq \lq bad\rq \rq} ones (whose corresponding values in the
background are nonzero). One should limit therefore the
analysis only to the \mbox{\lq \lq good\rq \rq} ones.

This same kind of behaviour seen
for the geometrical structure of the model also exists both for the kinematic
and dynamic quantities for the matter. Therefore the
\mbox{\lq \lq good\rq \rq} quantities which constitute the set of variables
with which we work should then be chosen from these particular scalars that
come from these three structures: geometric, kinematic and dynamic. Does that
mean that the present approach effectively avoids the gauge problem?

To answer this question affirmatively one should be able to exhibit a set of
\mbox{\lq \lq good\rq \rq} variables in such a way that its
corresponding dynamics is closed. That is, if we call ${\cal M}_{[A]}$ the
set of these variables, Einstein\rq s equations should provide the dynamics of
each element of ${\cal M}_{[A]}$, depending only on the background evolution
quantities (and, eventually, on other elements of ${\cal M}_{[A]}$). This
would exhaust the perturbation problem and we shall show in this paper that
this is indeed the case.

What should be learned from this discussion is that one should then
understand the gauge problem not as a basic difficulty on the perturbation
theory but just as a simple matter of asking a bad question\footnote{Let us
point out that some of the gauge dependent terms are particularly relevant,
$\delta\rho$ among these.}. One could imagine (what has been used a number of
times in the literature \cite{Hawking}, \cite{Olson}, \cite{MukhanovBrand})
that for FRW cosmology the perturbations of its main characteristics (the
energy density, $\delta\rho$ the scalar of curvature $\delta R$ and the
Hubble expansion factor $\delta\Theta$) would be natural quantities to be
considered as basic for the perturbation scheme. However, these are not
\mbox{\lq \lq good\rq \rq} scalars, since they are not zero in the
background
\footnote{However, as it will be seen in a next section, we can
construct 
associated \mbox{\lq \lq good\rq \rq} vector quantities in terms
of these scalars.}. We shall see in the next sections which scalars
replace these ones.

\subsection{SYNOPSIS}
\protect\label{synopsis}

In this paper we will deal only with observable quantities which
are associated to true perturbations of the cosmological FRW geometry as the
background. We are interested thus in the
variation of the electric part of the Weyl tensor\footnote{We will limit
ourselves here only to irrotational perturbation of the velocity
field. This implies that the corresponding magnetic part of Weyl tensor is
absent. To prove this statement one has to use eqs.(\ref{apb3}) and
(\ref{apb12}).}, along with the variation of shear and acceleration, since
these quantities are not induced by coordinate transformations.

We will analyse a certain set of
\mbox{\lq \lq good\rq \rq} scalar quantities which constitute a closed set,
i.e., one which provides a complete characterization of the perturbation
problem. Let us choose them as the Electric Weyl tensor $\delta E_{ij}$,
the shear $\delta \sigma_{ij}$ and the anisotropic stress $\delta \Pi_{ij}$
characterized by its corresponding magnitudes:
\begin{displaymath}
\sqrt{\delta E_{ij} \delta E^{ij}}
\end{displaymath}
\begin{displaymath}
\sqrt{\delta \Pi_{ij} \delta \Pi^{ij}}
\end{displaymath}
\begin{displaymath}
\sqrt{\delta \sigma_{ij} \delta \sigma^{ij}}
\end{displaymath}
for the geometry, the dynamics and the kinematics respectively. In the case of
perturbations allowing for vorticity we should add other invariants
containing the Magnetic Weyl tensor $\delta H_{ij}$. One should not consider
the restriction to the irrotational case as a limitation
of this method but instead as an attempt of making our approach clearer and
simpler in this paper. The application of this method to the case of vorticity
perturbation and gravitational waves will be dealt with in a forthcoming paper.

In Section \ref{definitions} we will present the complete set of definitions
and equations which will be needed in this paper. We include for completeness
the 14 Debever invariants and in Section \ref{fundamental} we will prove that
they are not suitable to produce a fundamental nucleus ${\cal M}_{[A]}$ from
which we would describe all our theory. Nevertheless one of the 14 invariants
will be included in ${\cal M}_{[A]}$. In this same section
we will characterize the set ${\cal M}_{[A]}$, whose dynamics will be given in
Section \ref{dynamics}. A comparison with previous gauge-invariant variables
is then established. For completeness the equations for some special remaining
gauge-dependent quantities ($\delta\rho$ and $\delta\Theta$) are also
exhibited. We shall see that, under very general circunstances, only the pair
of gauge-invariant quantities, $(E,\Sigma)$ (respectively the electric part
of Weyl tensor and the shear), constitutes a closed
dynamical system. This suggests the use of the gauge-invariant Hamiltonian
treatment for this problem. We shall then lay the foundation for such a
treatment, applied to these variables. This exhausts the total
problem of perturbation theory in the FRW background\footnote{However, if one
persists in asking questions about the evolution of intrinsically gauge-
dependent quantities (as for instance the perturbed density $\delta \rho$),
then a gauge must be obviously fixed. We would like to emphasize once again
that this does not constitute a drawback of the fundamental theory of
perturbation, once --- as we shall prove in the subsequent sections ---
it is possible, in order to solve this problem,
to deal with a complete closed system of differential
equations: this is precisely what will be done in this paper.}. In Section
\ref{lanczos} the Fierz-Lanczos potential is analysed
in the framework of FRW geometry. We exhibit the perturbation associated to
this tensor and its relationship with the fundamental kinematic quantities,
shear and acceleration. We end with Section \ref{conclusion}, in which some
general comments are given and future developments are sketched.

All equations we need (quasi-Maxwellian, the equations of constraint of matter
and the equations of evolution and constraints for the kinematical parameters
of a generic fluid) are presented in the Appendix.

\section{DEFINITIONS AND NOTATIONS}
\protect\label{definitions}

Greek indices run into the set $\{0, 1, 2, 3\}$. Latin indices run into the
set $\{1, 2, 3\}$. Weyl conformal tensor is defined by means of the
expression
\begin{equation}
W_{\alpha\beta\mu\nu} = R_{\alpha\beta\mu\nu} - M_{\alpha\beta\mu\nu}
+ \frac{1}{6} R g_{\alpha\beta\mu\nu}
\protect\label{d1}
\end{equation}
in which
\begin{equation}
g_{\alpha\beta\mu\nu} \equiv g_{\alpha\mu} g_{\beta\nu} - g_{\alpha\nu}
g_{\beta\mu}
\protect\label{d2}
\end{equation}
and
\begin{equation}
2 M_{\alpha\beta\mu\nu} = R_{\alpha\mu} g_{\beta\nu} + R_{\beta\nu}
g_{\alpha\mu} - R_{\alpha\nu} g_{\beta\mu} - R_{\beta\mu} g_{\alpha\nu}.
\protect\label{d3}
\end{equation}
We use the completely skewsymmetric Levi-Civita tensor
$\eta_{\alpha\beta\mu\nu}$ to perform the dual operation.The 10 algebraically
independent quantities of Weyl tensor can be separated in the
corresponding electric and magnetic parts, defined (by analogy with the
electromagnetic field) as:
\begin{equation}
E_{\alpha\beta} = - W_{\alpha\mu\beta\nu} V^{\mu} V^{\nu}
\protect\label{d4}
\end{equation}
\begin{equation}
H_{\alpha\beta} = - W^{\ast}_{\alpha\mu\beta\nu} V^{\mu} V^{\nu}.
\protect\label{d5}
\end{equation}
{}From the symmetry properties of Weyl tensor it follows that the dual
operation is independent on the pair in which it is applied.

These definitions yield that tensors $E_{\mu\nu}$ and $H_{\mu\nu}$ are
symmetric, traceless and belong to the tridimensional space, orthogonal to the
observer with 4-velocity $V^{\mu}$, that is:
\begin{equation}
\begin{array}{ll}
E_{\mu\nu} = E_{\nu\mu} \\ \\
E_{\mu\nu} V^{\mu} = 0 \\ \\
E_{\mu\nu} g^{\mu\nu} = 0,
\end{array}
\protect\label{d6}
\end{equation}
and
\begin{equation}
\begin{array}{ll}
H_{\mu\nu} = H_{\nu\mu} \\ \\
H_{\mu\nu} V^{\mu} = 0 \\ \\
H_{\mu\nu} g^{\mu\nu} = 0.
\end{array}
\protect\label{d7}
\end{equation}
The metric $g_{\mu\nu}$ and the vector $V^{\mu}$ (tangent to a timelike
congruence of curves $\Gamma$) induce a projector tensor
$h_{\mu\nu}$ which separates any tensor in terms of quantities defined along
$\Gamma$ plus quantities defined on the 3-dimensional space
${\cal H}$, orthogonal to $V^{\mu}$. The tensor $h_{\mu\nu}$, defined
in ${\cal H}$, is symmetric and a true projector, that is
\begin{equation}
h_{\mu\nu} h^{\nu\lambda} = {\delta_{\mu}}^{\lambda} - V_{\mu} \hspace{0.1cm}
V^{\lambda} = {h_{\mu}}^{\lambda}.
\protect\label{d8}
\end{equation}

FRW geometry is written in the standard Gaussian coordinate system as
\begin{equation}
ds^{2} = dt^{2} + g_{ij} dx^{i} dx^{j}
\protect\label{d9}
\end{equation}
in which $g_{ij} = - A^{2}(t) \gamma_{ij}(x^{k})$. The 3-dimensional geometry
has constant curvature and thus the corresponding Riemannian tensor
$^{(3)}R_{ijkl}$ can be written as
\begin{displaymath}
^{(3)}R_{ijkl} = K \gamma_{ijkl}.
\end{displaymath}
Covariant derivative in the
4-dimensional space-time will be denoted by the symbol (;) and the
3-dimensional derivative will be denoted by (${}_{\|}$).

The irreducible components of the covariant derivative of $V^{\mu}$ are given
in terms of the expansion scalar ($\Theta$), shear
($\sigma_{\alpha\beta}$), vorticity ($\omega_{\mu\nu}$) and acceleration
($a_{\alpha}$) by the standard definition:
\begin{equation}
V_{\alpha ;\beta} = \sigma_{\alpha\beta} + \frac{1}{3} \Theta h_{\alpha\beta}
+ \omega_{\alpha\beta} + a_{\alpha} V_{\beta}
\protect\label{d10}
\end{equation}
in which
\begin{equation}
\begin{array}{ll}
\sigma_{\alpha\beta} = \frac{1}{2} h^{\mu}_{(\alpha} h_{\beta)}^{\nu}
V_{\mu ;\nu} - \frac{1}{3} \Theta h_{\alpha\beta} \\ \\
\Theta = {V^{\alpha}}_{;\alpha} \\ \\
\omega_{\alpha\beta} = \frac{1}{2} h_{[\alpha}^{\mu} h_{\beta ]}^{\nu}
V_{\mu ;\nu} \\ \\
a_{\alpha} = V_{\alpha ;\beta} V^{\beta}
\end{array}
\protect\label{d11}
\end{equation}
We also define
\begin{equation}
\Theta_{\alpha\beta} \equiv \sigma_{\alpha\beta} + \frac{1}{3} \Theta
h_{\alpha\beta}.
\protect\label{d11b}
\end{equation}

Since the original Lifshitz paper \cite{Lifshitz} it has shown to be useful to
develop all perturbed quantities in the spherical harmonics basis. Once we
are limiting ourselves to irrotational perturbations, it is enough to our
purposes to take into account only the scalar $Q(x^{k})$ (with $\dot{Q} = 0$)
and its derived vector and tensor quantities. We have thus
\begin{equation}
\begin{array}{ll}
Q_{i} \equiv Q_{,i} \\ \\
Q_{ij} \equiv Q_{,i;j}
\end{array}
\protect\label{d12}
\end{equation}
where the scalar $Q$ obeys the eigenvalue equation defined in
the 3-dimensional background space by:
\begin{equation}
\bigtriangledown^{2}Q = mQ
\protect\label{d13}
\end{equation}
where $m$ is the wave number and
\begin{equation}
\bigtriangledown^{2}Q \equiv \gamma^{ik} Q_{,i\| k} = \gamma^{ik} Q_{,i;k}
\protect\label{d14}
\end{equation}
where the symbol $\bigtriangledown^{2}$ denotes the 3-dimensional Laplacian.
The traceless operator $\hat{Q}_{ij}$ is defined as
\begin{equation}
\hat{Q}_{ij} = \frac{1}{m} Q_{ij} - \frac{1}{3}Q \gamma_{ij}
\protect\label{d15}
\end{equation}
and the divergence of $\hat{Q}_{ij}$ is given by
\begin{equation}
{\hat{Q}^{ik}}_{  \| k} = 2 \left(\frac{1}{3} - \frac{K}{m}\right)
\hspace{0.1cm}Q^{i}.
\protect\label{d16}
\end{equation}
We remark that $Q$ is a 3-dimensional object; therefore indices are raised
with $\gamma^{ij}$, the 3-space metric.

In \cite{Debever} the complete 14 algebraically independent invariants
constructed with the curvature tensor were presented. Considering that we are
using an adimensional metric tensor, we can classify them with respect to
dimensionality as follows:

\vspace{0.5cm}
\begin{center}
\begin{tabular}{||c|c||}
\hline \hline
Dimensionality & Invariants \\ \hline
$L^{-2}$ & $I_{5}$ \\ \hline
$L^{-4}$ & $I_{1}$, $I_{3}$, $I_{6}$ \\ \hline
$L^{-6}$ & $I_{2}$, $I_{4}$, $I_{7}$, $I_{9}$, $I_{12}$ \\ \hline
$L^{-8}$ & $I_{8}$, $I_{10}$, $I_{13}$ \\ \hline
$L^{-10}$ & $I_{11}$, $I_{14}$ \\ \hline \hline
\end{tabular}
\end{center}
\vspace{0.5cm}

The expressions for these invariants are:
\begin{displaymath}
\begin{array}{ll}
I_{1} = W_{\alpha\beta\mu\nu} W^{\alpha\beta\mu\nu} \\ \\
I_{2} = {W_{\alpha\beta}}^{\rho\sigma} {W_{\rho\sigma}}^{  \mu\nu}
{W_{\mu\nu}}^{  \alpha\beta} \\ \\
I_{3} = W^{\alpha\beta\mu\nu} W^{\ast}_{\alpha\beta\mu\nu} \\ \\
I_{4} = W^{\alpha\beta\rho\sigma} {W_{\rho\sigma}}^{  \mu\nu}
W^{\ast}_{\mu\nu\alpha\beta} \\ \\
I_{5} = R \\ \\
I_{6} = C_{\mu\nu} C^{\mu\nu} \\ \\
I_{7} = C_{\alpha\beta} C^{\beta\mu} {C_{\mu}}^{\alpha} \\ \\
I_{8} = C_{\alpha\beta} C^{\beta\mu} C_{\mu\lambda} C^{\alpha\lambda} \\ \\
I_{9} = C_{\mu\nu} D^{\mu\nu} \\ \\
I_{10} = D_{\mu\nu} D^{\mu\nu} \\ \\
I_{11} = C_{\alpha\beta} D^{\beta\mu} {D_{\mu}}^{ \alpha} \\ \\
I_{12} = \tilde{D}_{\mu\nu} C^{\mu\nu} \\ \\
I_{13} = \tilde{D}_{\mu\nu} D^{\mu\nu} \\ \\
I_{14} = \tilde{D}_{\mu\nu} \tilde{D}^{\nu\alpha} {C^{\mu}}_{ \alpha}
\end{array}
\end{displaymath}
in which we used the following definitions:
\begin{equation}
\begin{array}{ll}
C_{\mu\nu} \equiv R_{\mu\nu} - \frac{1}{4} R g_{\mu\nu} \\ \\
D_{\mu\nu} \equiv W_{\mu\alpha\nu\beta} C^{\alpha\beta} \\ \\
\tilde{D}_{\mu\nu} \equiv W^{\ast}_{\mu\alpha\nu\beta}
C^{\alpha\beta}.
\end{array}
\protect\label{d17}
\end{equation}

\section{FUNDAMENTAL PERTURBATIONS OF FRW UNIVERSE}
\protect\label{fundamental}

As we observed in the previous section, a complete examination of the
perturbation theory should naturally include the analysis of the evolution of
the Debever metric invariants associated to FRW geometry.

The only non identically zero invariants of FRW geometry are given by
\begin{displaymath}
\begin{array}{ll}
I_{5} = (1 - 3\lambda) \rho \\ \\
I_{6} = \frac{3}{4} (1 + \lambda)^{2} \rho^{2} \\ \\
I_{7} = - \frac{3}{8} (1 + \lambda)^{3} \rho^{3} \\ \\
I_{8} = \frac{21}{64} (1 + \lambda)^{4} \rho^{4}
\end{array}
\end{displaymath}
in which we used Einstein\rq s equations
\begin{displaymath}
R_{\mu\nu} - \frac{1}{2} R g_{\mu\nu} = - T_{\mu\nu}
\end{displaymath}
and the stress-energy tensor is that of a perfect fluid
\begin{displaymath}
T_{\mu\nu} = (1 + \lambda) \rho V_{\mu} V_{\nu} - \lambda \rho g_{\mu\nu}.
\end{displaymath}

If we restrict ourselves to the linear perturbation theory, the only
invariants which have non identically zero linear perturbation terms
are $I_{5}$, $I_{6}$, $I_{7}$, $I_{8}$,
$I_{9}$ and $I_{12}$. Among these the first four are nonzero
in the background and the latter two are zero, since the geometry is
conformally flat. This could lead to the conclusion that $I_{9}$ and
$I_{12}$ are the \mbox{\lq \lq good\rq \rq} scalars to be examined. However,
a direct calculation shows that the latter two invariants have zero
linear perturbation. Indeed, it follows from FRW geometry that the
perturbation of $I_{9}$ reduces to
\begin{displaymath}
\delta I_{9} = C^{\mu\nu} C^{\alpha\beta} \delta W_{\mu\alpha\nu\beta}.
\end{displaymath}
Then (due to the fact that Weyl tensor is trace-free) the above quantity
vanishes identically. This result depends of course on
the fact that the source of the background geometry is given by a perfect
fluid. In effect we have in this case
\begin{displaymath}
\delta I_{9} =  (\rho + p)^{2} \left(V^{\mu} V^{\nu} -
\frac{1}{4} g^{\mu\nu} \right)
\left(V^{\alpha} V^{\beta} - \frac{1}{4} g^{\alpha\beta} \right)
\delta W_{\mu\alpha\nu\beta}
\end{displaymath}
which is zero. For the same reasoning $\delta I_{12}$, given by
\begin{displaymath}
\delta I_{12} = C^{\mu\nu} C^{\alpha\beta} \delta
W^{\ast}_{\mu\alpha\nu\beta}
\end{displaymath}
also vanishes.

The corresponding perturbations for the remaining invariants are given by
\begin{displaymath}
\begin{array}{ll}
\delta I_{5} = (1 - 3\lambda)\delta\rho \\ \\
\delta I_{6} = \frac{3}{2} (1 + \lambda)^{2}\rho \hspace{0.1cm}\delta\rho \\ \\
\delta I_{7} = - \frac{9}{8} (1 + \lambda)^{3} \rho^{2} \hspace{0.1cm}
\delta\rho \\ \\
\delta I_{8} = \frac{21}{16} (1 + \lambda)^{4} \rho^{3} \hspace{0.1cm}
\delta\rho.
\end{array}
\end{displaymath}

It follows from these results that the perturbations of these quantities
are algebraically related\footnote{One can write these invariants
in a pure geometrical way without using Einstein\rq s equations. This
does not modify our argument.}.
Besides, once all these scalars have a non-zero
background value, they do not belong to the minimum set of good
quantities that we are searching for.

Corresponding difficulties occur for the standard kinematical and dynamical
variables, that is, the expansion parameter $\Theta$ and the density of energy
$\rho$ suffer from the same disease.

This is thus the bad choice for the basic variables which we should avoid.
Let us now turn our attention to which good variables
should be considered as the fundamental ones.

\subsection{GEOMETRIC PERTURBATION}
\protect\label{geometric}

{}From the previous section it follows that
\begin{displaymath}
\sqrt{\delta E_{ij} \delta E^{ij}}
\end{displaymath}
is the only quantity that characterizes without ambiguity a true
perturbation of the Debever invariants\footnote{This is a consequence of
the vanishing of the perturbation of the magnetic part of
Weyl tensor(cf.above).}. We need thus to
consider only the perturbed $E_{ij}$ since, as we shall see,
any other metric quantity does not belong to the \mbox{\lq \lq good\rq \rq}
basic nucleus needed for a complete
knowledge of the true perturbations. We then set the expansion of this
tensor in terms of the spherical harmonic basis
\begin{equation}
\delta E_{ij} = E(t) \hspace{0.1cm}\hat{Q}_{ij}(x^{k}).
\protect\label{g1}
\end{equation}
Thus $E(t)$ is the geometric quantity whose dynamics we are looking for.

\subsection{KINEMATICAL PERTURBATIONS}
\protect\label{kinematical}

We restrict our considerations only to linear perturbation terms. The
normalization of the 4-velocity yields that the variation of the time
component of the perturbed velocity is related to the variation of the
(0-0) component of the metric tensor, that is:
\begin{equation}
\delta V_{0} = \frac{1}{2} \delta g_{00}.
\protect\label{k1}
\end{equation}
The corresponding contravariant quantities are related as follows:
\begin{equation}
\delta V^{0} = \frac{1}{2} \delta g^{00} = - \delta V_{0}.
\protect\label{k2}
\end{equation}

The expansion of the perturbations of the 4-velocity in terms of the
spherical harmonic basis is\footnote{The vorticity
is of course zero, since we are limiting ourselves to the irrotational case.}
\begin{equation}
\begin{array}{ll}
\delta V_{0} = \frac{1}{2} \beta (t) \hspace{0.1cm}Q(x^{i})
+ \frac{1}{2} Y(t) \\ \\
\delta V_{k} = V(t) \hspace{0.1cm}Q_{k}(x^{i}).
\end{array}
\protect\label{k3}
\end{equation}
For the acceleration we set
\begin{equation}
\delta a_{k} = \Psi (t) \hspace{0.1cm}Q_{k}(x^{i}).
\protect\label{k4}
\end{equation}
For the shear
\begin{equation}
\delta \sigma_{ij} = \Sigma (t) \hspace{0.1cm}\hat{Q}_{ij}(x^{k})
\protect\label{k5}
\end{equation}
and for the expansion we set
\begin{equation}
\delta \Theta = H(t) \hspace{0.1cm}Q(x^{i}) + Z(t)
\protect\label{k6}
\end{equation}
where $Y(t)$ and $Z(t)$ are homogeneous terms that are {\bf not} true
perturbations.

Let us point out that, once we are limiting ourselves to the analysis of true
perturbed quantities, the important kinematical variable
whose dynamics we need to examine is only $\Sigma(t)$,
since the other gauge-invariant quantity $\Psi$ is a function of $\Sigma$ (and
$E$), as we shall see ($\beta$ is just a matter of choice of the coordinate
system).

\subsection{MATTER PERTURBATION}
\protect\label{matter}

Since we are considering a background geometry in which there is a state
equation relating the pressure and the energy density, i.e. $p = \lambda
\rho$, we will consider the standard procedure that accepts the
preservation of this state equation under arbitrary perturbations. Besides,
our frame is such that there is no heat flux. Thus the general form of the
perturbed energy-momentum tensor is given by
\begin{equation}
\delta T_{\mu\nu} = (1 + \lambda) \hspace{0.1cm}\delta (\rho V_{\mu} V_{\nu})
- \lambda \delta (\rho g_{\mu\nu}) + \delta \Pi_{\mu\nu}.
\protect\label{m1}
\end{equation}

We write $\delta \rho$ in terms of the scalar basis as:
\begin{equation}
\delta\rho = N(t) \hspace{0.1cm}Q(x^{i}) + \mu (t)
\protect\label{m2}
\end{equation}
in which the homogeneous term $\mu (t)$ is not a true
perturbation\footnote{We will set $Y = Z = \mu = 0$, since
these homogeneous terms are just a matter of choice of the coordinate system.
Nevertheless we are not interested in examining pure gauge quantities such as
$Y$, $Z$ and $\mu$.}.

According to causal thermodynamics the evolution equation of the anisotropic
pressure is related to the shear through \cite{Israel}
\begin{equation}
\tau\dot{\Pi}_{ij} + \Pi_{ij} = \xi\sigma_{ij}
\protect\label{extra1}
\end{equation}
in which $\tau$ is the relaxation parameter and $\xi$ is the viscosity
parameter. For simplicity of this present treatment we will limit ourselves
to the case in which $\tau$ can be neglected and $\xi$ is a constant
\footnote{In the general case $\xi$ and $\tau$ are functions of the
equilibrium variables, for instance $\rho$ and the temperature $T$ and, since
both variations $\delta\Pi_{ij}$ and $\delta\sigma_{ij}$ are expanded in
terms of the traceless tensor $\hat{Q}_{ij}$, it follows that
the above
relation does not restrain the kind of fluid we are examining. However, if we
consider $\xi$ as time-dependent, the quantity $\delta\Pi_{ij}$ must be
included in the fundamental set ${\cal M}_{[A]}$.}; eq.(\ref{extra1}) then
gives
\begin{equation}
\Pi_{ij} = \xi\sigma_{ij}
\protect\label{extra2}
\end{equation}
and the associated perturbed equation is:
\begin{equation}
\delta\Pi_{ij} = \xi \hspace{0.1cm}\delta\sigma_{ij}.
\protect\label{m3}
\end{equation}

Following the same reasoning as before, $\delta \Pi_{ij}$ is the matter
quantity that should
enter in the complete system of differential equations which describes
the perturbation evolution. One should also be interested in the
dynamics of $\delta\rho$ although it is not a fundamental part of the
basic system of equations. We will examine its evolution later on.

The \mbox{\lq \lq good\rq \rq} set ${\cal M}_{[A]}$ has therefore three
elements: $\delta E_{ij}$, $\delta\sigma_{ij}$ and $\delta\Pi_{ij}$. But,
since $\delta\Pi_{ij}$ is written in terms of $\delta\sigma_{ij}$, the
set ${\cal M}_{[A]}$ which will be considered reduces to:
\begin{displaymath}
{\cal M}_{[A]} = \{\delta E_{ij}, \delta\sigma_{ij}\}.
\end{displaymath}
So much for definitions. Let us then turn to the analysis of the dynamics.

\section{DYNAMICS}
\protect\label{dynamics}

In this section we will show that $E(t)$ and $\Sigma (t)$ constitute the
fundamental pair of variables in terms of which all the dynamics for
the perturbed FRW geometry is given, that is,
\mbox{${\cal M}_{[A]} = \{E(t), \Sigma (t)\}$} is the minimal closed
set of observables in the perturbation theory of FRW which characterizes and
determines completely the spectrum of perturbations. Indeed, the evolution
equations for these two quantities (which come from Einstein\rq s equations)
generate a dynamical system involving only $E$ and $\Sigma$ (and background
quantities) which, when solved, contains all the necessary information for a
complete description of all remaining perturbed quantities of FRW
geometry. Such a conclusion does not seem to have been noticed in the past.

We remark that we will limit ourselves only to the examination of the
perturbed quantities that are relevant for the complete
knowledge of the system. These equations are the quasi-Maxwellian equations
of gravitation and the evolution
equations for the kinematical quantities. In \cite{Vishniac} and
\cite{NovelloSalim} this system of equations was presented and analysed; we
will list them in the Appendix for completeness.

\subsection{THE PERTURBED EQUATION FOR THE SHEAR}
\protect\label{shear}

The perturbed equation for the shear eq.(\ref{apb18}) is written as:
\begin{eqnarray}
{h_{\alpha}}^{\mu} \hspace{0.1cm} {h_{\beta}}^{\nu} \hspace{0.1cm}(\delta
\sigma_{\mu\nu})^{\bullet} & + & \frac{2}{3} \Theta
\hspace{0.1cm}\delta\sigma_{\alpha
\beta} + \frac{1}{3} h_{\alpha\beta} \hspace{0.1cm}\delta
{a^{\lambda}}_{;\lambda} \nonumber \\
& - & \frac{1}{2}{h_{\alpha}}^{\mu} \hspace{0.1cm}{h_{\beta}}^{\nu}
\hspace{0.1cm}[\delta a_{\mu ;\nu} + \delta a_{\nu ;\mu}] \nonumber \\
& = & \delta M_{\alpha\beta}
\protect\label{s1}
\end{eqnarray}
where
\begin{equation}
M_{\alpha\beta} \equiv R_{\alpha\mu\beta\nu} \hspace{0.1cm}V^{\mu} V^{\nu} -
\frac{1}{3} R_{\mu\nu} \hspace{0.1cm}V^{\mu}V^{\nu} \hspace{0.1cm}h_{\alpha
\beta}.
\protect\label{s2}
\end{equation}

Using the above spherical harmonics expansion and eq.(\ref{m3}),
eq.(\ref{s1}) reduces to:
\begin{equation}
\dot{\Sigma} = -E - \frac{1}{2}\xi \hspace{0.1cm}\Sigma + m \hspace{0.1cm}
\Psi.
\protect\label{s3}
\end{equation}

\subsection{THE PERTURBED EQUATION FOR $E_{ij}$}
\protect\label{eletrica}

The perturbed equation for the electric part of the Weyl tensor
is given in the Appendix. Using the above spherical harmonics
expansion and eq.(\ref{m3}) one obtains:
\begin{eqnarray}
\dot{E} & = & - \frac{(1 + \lambda)}{2} \rho \hspace{0.1cm}\Sigma -
\left(\frac{\Theta}{3} + \frac{\xi}{2}\right) \hspace{0.1cm}E \nonumber \\
& - & \frac{\xi}{2} \hspace{0.1cm}\left(\frac{\xi}{2} +
\frac{\Theta}{3}\right) \hspace{0.1cm} \Sigma + \frac{m}{2} \hspace{0.1cm}\xi
\hspace{0.1cm}\Psi.
\protect\label{e2}
\end{eqnarray}
This suggests that $E$ and $\Sigma$ may be considered as canonically
conjugated variables. We shall see later on that this is indeed the case.

Equations (\ref{s3}) and (\ref{e2}) contain three variables: $E$, $\Sigma$ and
$\Psi$. We will now show that using the conservation law for the matter we
can eliminate $\Psi$ in all cases, except when $(1 + \lambda) = 0$. We will
return to this particular (vacuum) case in a later section.

The proof is the following. Projecting the conservation equation of the
energy-momentum tensor in the 3-space, that is
\begin{equation}
{T^{\mu\nu}}_{;\nu} {h_{\mu}}^{\lambda} = 0
\protect\label{e3}
\end{equation}
and using the perturbed quantities this equation gives:
\begin{equation}
(1 + \lambda) \rho \hspace{0.1cm}\delta a_{k} - \lambda (\delta \rho)_{,k} +
\lambda \dot{\rho} \hspace{0.1cm}\delta V_{k} + \delta {{\Pi_{k}}^{i}}_{;i}
= 0.
\protect\label{e4}
\end{equation}

Using the decomposition in the spherical harmonics basis we obtain
\begin{equation}
(1 + \lambda) \rho \Psi = \lambda [N - \dot{\rho} V] +
2\xi \left(\frac{1}{3} - \frac{K}{m}\right) \hspace{0.1cm}A^{-2}
\hspace{0.1cm}\Sigma.
\protect\label{e5}
\end{equation}

Now comes a remarkable result: the right hand side of eq.(\ref{e5}) can be
expressed in terms of the variables $E$ and $\Sigma$ only (since we are
analysing here the case where $(1 + \lambda)$ does not vanish).
Indeed, from the equation of divergence of the electric tensor (see Appendix),
we find
\begin{equation}
N - \dot{\rho} V = \left(1 - \frac{3K}{m}\right) \hspace{0.1cm}\xi
\hspace{0.1cm}\Sigma A^{-2} - 2 \left(1 - \frac{3K}{m}\right) \hspace{0.1cm}
A^{-2} \hspace{0.1cm}E.
\protect\label{e6}
\end{equation}

Combining these two equations we find that $\Psi$ is given in terms of
the background quantities and the basic perturbed terms $E$ and $\Sigma$:
\begin{equation}
(1 + \lambda) \hspace{0.1cm}\rho \hspace{0.1cm}\Psi = 2
\left(1 - \frac{3K}{m}\right) \hspace{0.1cm}A^{-2} \hspace{0.1cm}
[ -\lambda E + \frac{1}{2} \hspace{0.1cm}\lambda \hspace{0.1cm}\xi
\hspace{0.1cm}\Sigma + \frac{1}{3} \hspace{0.1cm}\xi \hspace{0.1cm}\Sigma].
\protect\label{e7}
\end{equation}

Thus the whole set of perturbed equations reduces, for the variables
$E$ and $\Sigma$, to a time-dependent dynamical system:
\begin{equation}
\begin{array}{ll}
\dot{\Sigma} = F_{1}(\Sigma, E) \\ \\
\dot{E} = F_{2}(\Sigma, E)
\end{array}
\protect\label{e8}
\end{equation}
with
\begin{displaymath}
F_{1} \equiv -E - \frac{1}{2}\xi \hspace{0.1cm}\Sigma + m \hspace{0.1cm}
\Psi
\end{displaymath}
and
\begin{eqnarray}
F_{2} & \equiv & - \left(\frac{1}{3}\Theta + \frac{1}{2}\xi\right)
\hspace{0.1cm}E \nonumber \\
& - & \left(\frac{1}{4}\xi^{2} + \frac{(1 + \lambda )}{2}\rho + \frac{1}{6}
\xi\Theta\right) \hspace{0.1cm}\Sigma \nonumber \\
& + & \frac{m}{2} \hspace{0.1cm}\xi \hspace{0.1cm}\Psi \nonumber
\end{eqnarray}
in which $\Psi$ is given in terms of $E$ and $\Sigma$ by eq.(\ref{e7}).

\subsection{COMPARISON WITH PREVIOUS GAUGE-INVARIANT VARIABLES}
\protect\label{previous}

FRW cosmology is characterized by the homogeneity of the fundamental
variables that specify its kinematics (the expansion factor $\Theta$),
its dynamics (the energy density $\rho$) and its associated geometry
(the scalar of curvature $R$). This means that these three quantities
depend only on the global time $t$, characterized by the hypersurfaces
of homogeneity. We can thus use this fact to define in a trivial way
3-tensor associated quantities, which vanish in this geometry, and
look for its corresponding non-identically vanishing perturbation. The
simplest way to do this is just to let $U$ be a homogeneous
variable (in the present case, it can be any one of the quantities $\rho$,
$\Theta$ or $R$), that is $U = U(t)$. Then use the 3-gradient operator
$^{(3)}\bigtriangledown_{\mu}$ defined by
\begin{equation}
^{(3)}\bigtriangledown_{\mu} \equiv
{h_{\mu}}^{\lambda} \hspace{0.1cm}\bigtriangledown_{\lambda}
\protect\label{prev1}
\end{equation}
to produce the desired associated variable
\begin{equation}
U_{\mu} = {h_{\mu}}^{\lambda} \hspace{0.1cm}\bigtriangledown_{\lambda} U.
\protect\label{prev2}
\end{equation}

In \cite{Ellis et al} these quantities were discussed and
its associated evolution
analysed. In the present section we will exhibit the relation of these
variables to our fundamental ones. We shall see that under the conditions of
our analysis\footnote{We remind the reader that we restrain here our
examination to irrotational perturbation. The formulas which we obtain are
thus simpler. However the method of our analysis is not restrictive and
the study of generic cases can be obtained through the same
lines.} these quantities are functionals of our basic variables ($E$
and $\Sigma$) and the background ones.

\subsection{THE MATTER VARIABLE $\chi_{i}$}
\protect\label{matter variable}

It seems useful to define the fractional gradient of the energy density
$\chi_{\alpha}$ as \cite{Ellis et al}
\begin{equation}
\chi_{\alpha} \equiv \frac{1}{\rho} \hspace{0.1cm}
^{(3)}\bigtriangledown_{\alpha} \hspace{0.1cm}\rho.
\protect\label{1234}
\end{equation}

Such quantity $\chi_{\alpha}$ is nothing but a combination of the
acceleration and the divergence of the anisotropic stress. Indeed, from
eq.(\ref{e4}) it follows (in the frame in which there is no heat flux)
\begin{equation}
\delta \chi_{i} = \frac{(1 + \lambda)}{\lambda} \hspace{0.1cm}
\delta a_{i} + \frac{1}{\lambda\rho} \hspace{0.1cm}
\delta{{\Pi_{i}}^{\beta}}_{;\beta}
\protect\label{1235}
\end{equation}

{}From what we have learned above it follows that this quantity can be
reduced to a functional of the basic quantities of perturbation,
that is $\Sigma$ and $E$, yielding
\begin{equation}
\delta \chi_{i} = -2 \hspace{0.1cm}\left(1 - \frac{3K}{m}\right) \hspace{0.1cm}
\frac{1}{\rho A^{2}} \hspace{0.1cm}\left(E - \frac{\xi}{2} \Sigma\right)
\hspace{0.1cm}Q_{i}.
\protect\label{1236}
\end{equation}

\subsection{THE KINEMATICAL VARIABLE $\eta_{i}$}
\protect\label{kinematical variable}

The only non-vanishing quantity of the kinematics of the cosmic
background fluid is the (Hubble) expansion factor $\Theta$. This allows us to
define the quantity $\eta_{\alpha}$ as:
\begin{equation}
\eta_{\alpha} = {h_{\alpha}}^{\beta} \hspace{0.1cm}\Theta_{,\beta}.
\protect\label{1237}
\end{equation}
Using the constraint relation eq.(\ref{apb10}) we can relate this quantity
to the basic ones:
\begin{equation}
\delta \eta_{i} = -\frac{\Sigma}{A^{2}} \hspace{0.1cm}\left(1 - \frac{3K}{m}
\right) \hspace{0.1cm}Q_{i}.
\protect\label{1238}
\end{equation}

\subsection{THE GEOMETRICAL VARIABLE $\tau$}
\protect\label{the geometrical variable}

We can choose the scalar of curvature $R$ which depends only on the
cosmical time $t$ like $\rho$ and $\Theta$ to be the
$U$-geometrical variable.
However it seems more appealing to use a combined expression $\tau$
involving $R$, $\rho$ and $\Theta$ given by
\begin{equation}
\tau = R + (1 + 3\lambda) \hspace{0.1cm}\rho - \frac{2}{3} \hspace{0.1cm}
\Theta^{2}.
\protect\label{1239}
\end{equation}
In the unperturbed FRW background this quantity is defined in terms
of the curvature scalar of the 3-dimensional space and the scale factor
$A(t)$:
\begin{displaymath}
\frac{^{(3)}R}{A^{2}}.
\end{displaymath}
We define then the new associated variable $\tau_{\alpha}$ as
\begin{equation}
\tau_{\alpha} = {h_{\alpha}}^{\beta} \hspace{0.1cm}\tau_{,\beta}.
\protect\label{1311}
\end{equation}
This quantity $\tau_{\alpha}$ vanishes in the background. Its perturbation can
be written in terms of the previous variations, since Einstein\rq s
equations give
\begin{displaymath}
\tau = 2 \hspace{0.1cm}\left(\rho - \frac{1}{3} \hspace{0.1cm}
\Theta^{2}\right).
\end{displaymath}

We can thus, without any information loss, limit all our
analysis to the fundamental variables. Nevertheless, just for
completeness, let us exhibit the evolution equations for some
gauge-dependent variables.

\subsection{PERTURBED EQUATIONS FOR $\rho$ AND $\Theta$}
\protect\label{rhoetheta}

{}From eq.(\ref{apb23}) and using the decomposition of the perturbed energy
density in the scalar basis (eq.(\ref{m2})) we obtain the equation of
evolution for $\delta\rho$ as:
\begin{equation}
\dot{N} - \frac{1}{2} \hspace{0.1cm}\beta \hspace{0.1cm}\dot{\rho} +
(1 + \lambda) \hspace{0.1cm}\Theta \hspace{0.1cm}N + (1 + \lambda)
\hspace{0.1cm}\rho \hspace{0.1cm}H = 0.
\protect\label{rt1}
\end{equation}

Applying the same procedure for the perturbed Raychaudhuri equation
(eq.(\ref{apb20})) and using the decomposition eq.(\ref{k6}) we obtain
\begin{equation}
\dot{H} - \frac{1}{2}\hspace{0.1cm}\beta \hspace{0.1cm}\dot{\Theta} +
\frac{2}{3}\hspace{0.1cm}\Theta \hspace{0.1cm}H +
\frac{m}{A^{2}} \hspace{0.1cm}\Psi + \frac{(1 + 3\lambda)}{2} \hspace{0.1cm}N
= 0.
\protect\label{rt2}
\end{equation}

To solve these two equations we need to fix the gauge ($\beta (t)$) and to
use the values for $E$ and $\Sigma$ which were obtained from the fundamental
closed system found in the previous section (eqs.(\ref{e8})). All the
remaining geometrical and kinematical quantities can be likewise obtained.
This exhausts completely our analysis of the irrotational perturbations of
FRW universe.

\subsection{THE SINGULAR CASE $(1 + \lambda) = 0$: THE PERTURBATIONS OF DE
SITTER UNIVERSE}
\protect\label{deSitter}

We have seen that all the system of reduction to the variables $\Sigma$ and
$E$ was based on the possibility of writing the acceleration in terms of
$E$ and $\Sigma$. This was possible in all cases, except in the special one
in which $(1 + \lambda) = 0$. Although no known fluid exists with such
negative pressure, the fact that the vacuum admits such an interpretation has
led to the identification of the cosmological constant with this fluid. It is
therefore worthwhile to examine this case in the same way as it was done for
the previous sections.

At this point it must be remarked that, contrarily from all the previously
studied cases, perturbations of this fluid must
necessarily contain contributions which come from the heat flux or the
anisotropic pressure. Indeed, if we take both of these quantities as
vanishing, then the set of perturbed equations implies that all equations
are trivially satisfied, since all perturbative quantities vanish, except for
the cases where $\delta p = \overline{\lambda} \hspace{0.1cm}\delta\rho$,
with $\overline{\lambda} = 0$, and $\lambda + 1 = 0$. We will
analyse these cases below.

When $\delta p = \overline{\lambda} \hspace{0.1cm}\delta\rho$,
for $\overline{\lambda} = 0$, the system is stable. Indeed, we obtain for
the electric part of Weyl tensor, in the case that $\Theta$ is constant in the
background, the following expression:
\begin{displaymath}
E(t) = E_{0} \hspace{0.1cm}e^{- \frac{\Theta}{3} \hspace{0.1cm}t}.
\end{displaymath}

The other case of interest is the one in which the condition
\mbox{$(1 + \lambda) = 0$} is
preserved throughout the perturbation. Looking at eq.(\ref{rt1}) it follows
that, from the fact that $\dot{\rho} = 0$ and reminding the reader that
$(1 + \lambda) = 0$, temporal variation of the energy density exists only if
we take into account the perturbed fluid with heat flux. We then write
\begin{displaymath}
q_{i} = q(t) \hspace{0.1cm} Q_{i}(x^{k}).
\end{displaymath}
Equation (\ref{apb23}) gives
\begin{equation}
\dot{N} = \frac{m}{A^{2}} \hspace{0.1cm}q.
\protect\label{dS1}
\end{equation}

The projected conserved equation gives (see eq.(\ref{apb24})):
\begin{equation}
\dot{q} + \Theta \hspace{0.1cm}q + N = \frac{2\xi}{3A^{2}} \hspace{0.1cm}
\left(1 - \frac{3K}{m}\right) \hspace{0.1cm}\Sigma.
\protect\label{dS2}
\end{equation}

The evolution equation for the electric part of Weyl tensor gives:
\begin{equation}
\dot{{\cal S}} + \frac{\Theta}{3} \hspace{0.1cm}{\cal S} = - \frac{m}{2}
\hspace{0.1cm}q
\protect\label{dS3}
\end{equation}
in which we used the definition
\begin{displaymath}
{\cal S} \equiv E - \frac{1}{2} \hspace{0.1cm}\xi \hspace{0.1cm}\Sigma.
\end{displaymath}

Finally, from the equation that gives the divergence of $E_{ij}$, we have the
constraint
\begin{equation}
\frac{2}{A^{2}} \hspace{0.1cm}\left(1 - \frac{3K}{m}\right)
\hspace{0.1cm}{\cal S} = - \left(N + \Theta \hspace{0.1cm}q\right).
\protect\label{dS4}
\end{equation}

The evolution equation for the shear provides the value of the
acceleration $\Psi$. Equations (\ref{dS2})-(\ref{dS4}) constitute thus a
complete system for the variables $E$, $\Sigma$ and $q$. This completes the
general explicitly gauge-invariant scheme that we presented here even in the
singular case $(1 + \lambda) = 0$.
Notwithstanding, just as an additional comment, it would be interesting
to consider the perturbation scheme in the framework of Lanczos potential.
This will be done in a later section.

\subsection{HAMILTONIAN TREATMENT}
\protect\label{Hamiltonian}

The examination of the perturbations in FRW cosmology, which we analysed
above, admits a Hamiltonian formulation that is worth considering here
\cite{Grishchuk}. In this vein, the variables $E$ and $\Sigma$, analysed
in the previous section, are the ones that must be employed to obtain such a
formulation. From the evolution equations for $\Sigma$ and $E$ (eq.(\ref{e8}))
it follows that they are not canonically conjugated for arbitrary geometries
of the background.

The natural step would be to define canonically conjugated variables $Q$ and
$P$ as a linear functional of $\Sigma$ and $E$ as\footnote{The
attentive reader should notice that in this subsection the quantity $Q$ shall
not be confused with the previous scalar basis.}:
\begin{equation}
\left[\begin{array}{cc}
Q \\
P
\end{array}\right] = \left[\begin{array}{ccc}
\alpha & \eta \\
\delta & \beta
\end{array}\right] \hspace{0.2cm}\left[\begin{array}{cc}
\Sigma \\
E
\end{array}\right].
\protect\label{h1A}
\end{equation}
It should be expected that functionals of the background geometry would
appear in the construction of the canonical variables in the functions
$\alpha$, $\beta$, $\eta$ and $\delta$. It seems worth to remark that this
matrix is univocally defined up to canonical transformations. We can thus use
this fact to choose $\eta$ and $\delta$ as
zero; we shall use this choice in order to simplify our analysis.

The Hamiltonian ${\cal H}$ which provides the dynamics of the pair $(Q,P)$ is
obtained from the evolution equations of $E$ and $\Sigma$ (\ref{e8}). The
condition for the existence of such a Hamiltonian is given by the equation
\begin{eqnarray}
\frac{\dot{\alpha}}{\alpha} & + & \frac{\dot{\beta}}{\beta} - \xi -
\frac{1}{3} \hspace{0.1cm}\Theta \nonumber \\
& + & \frac{2m\xi}{3(1 + \lambda) \hspace{0.1cm}\rho \hspace{0.1cm}A^{2}}
\hspace{0.1cm}\left(1 - \frac{3K}{m}\right) = 0.
\protect\label{h2A}
\end{eqnarray}

It then follows that the Hamiltonian which provides the dynamics of our
problem takes the form
\begin{equation}
{\cal H} = \frac{h_{1}}{2} \hspace{0.1cm}Q^{2} + \frac{h_{2}}{2}
\hspace{0.1cm}P^{2} +
2 \hspace{0.1cm}h_{3} \hspace{0.1cm}P \hspace{0.1cm}Q
\protect\label{h3A}
\end{equation}
where $h_{1}$, $h_{2}$ and $h_{3}$ are defined as
\begin{eqnarray}
h_{1} \equiv \frac{\beta}{\alpha} & \{ & \frac{(1 + \lambda)}{2}
\hspace{0.1cm}\rho +
\frac{\xi}{2} \hspace{0.1cm}\left(\frac{\xi}{2} + \frac{\Theta}{3}\right)
\nonumber \\
& - & \frac{m \hspace{0.1cm}\xi^{2}}{(1 + \lambda) \hspace{0.1cm}\rho A^{2}}
\hspace{0.1cm}\left(1 - \frac{3K}{m}\right) \hspace{0.1cm}\left(
\frac{\lambda}{2} + \frac{1}{3}\right) \}
\protect\label{h4A}
\end{eqnarray}
\begin{equation}
h_{2} \equiv - \frac{\alpha}{\beta} \{1 + \frac{2m\lambda}{(1 + \lambda)
\hspace{0.1cm}\rho
A^{2}} \hspace{0.1cm}\left(1 - \frac{3K}{m}\right)\}
\protect\label{h5A}
\end{equation}
\begin{equation}
h_{3} \equiv \frac{\xi}{4} \hspace{0.1cm}\{1 + \frac{2m\lambda}{(1 + \lambda)
\hspace{0.1cm}\rho A^{2}} \hspace{0.1cm}\left(1 - \frac{3K}{m}\right)^{2}\}.
\protect\label{h6A}
\end{equation}

Let us consider the case in which $\xi = 0$, that is, there is no anisotropic
pressure. The case where $\xi$ does not vanish presents some interesting
peculiarities which will be left to a forthcoming paper.

We will choose $\beta = A$ and take $\alpha$ as given by eq.(\ref{h2A}). We
then define the canonical variables $Q$ and $P$ by setting
\begin{displaymath}
\begin{array}{ll}
Q = \Sigma \\ \\
P = A \hspace{0.1cm}E.
\end{array}
\end{displaymath}

It then follows that ${\cal H}$ is given by
\begin{equation}
{\cal H} =  - \Delta^{2}(t) \hspace{0.1cm}P^{2} + \gamma^{2}(t)
\hspace{0.1cm}Q^{2}
\protect\label{h1}
\end{equation}
where $\gamma (t)$ and $\Delta (t)$ are given in terms of the energy density
of the background $\rho$, the scale factor $A(t)$ and the wave number $m$ as:
\begin{equation}
\begin{array}{ll}
\gamma^{2}(t) \equiv \left(\frac{(1 + \lambda)}{4}\right) \hspace{0.1cm}\rho
\hspace{0.1cm}A \\ \\
\Delta^{2}(t) \equiv \frac{1}{2A} \left(1 +
\frac{2m\lambda}{(1 + \lambda)\rho \hspace{0.1cm}A^{2}} \hspace{0.1cm}
\left(1 - \frac{3K}{m}\right)\right).
\end{array}
\protect\label{h2}
\end{equation}

Let us make two comments here: first of all, the fact that the system is not
conservative (which means $\dot{{\cal H}}$ is not zero) is a consequence of
the fact that the ground state of this theory ($Q = P = 0$) corresponds not to
Minkowskii flat space-time but to FRW expanding universe. The second remark
is that the same applies to the non-positivity of the Hamiltonian; this is
also a consequence of the non-vanishing of the curvature of the fundamental
state. The system which we are analysing is not closed and so momentum and
energy can be pumped from the background.

We notice that the Hamiltonian structure obtained in terms of the variables
$E$ and $\Sigma$ is completely gauge-invariant and, as such, deserves an
ulterior analysis, which we will make elsewhere. We would like only to
exhibit an example where this pumping effect can be easily recognized; this
will be achieved by applying the Hamiltonian treatment to a static model of
the universe.

\subsubsection{EINSTEIN\rq S STATIC UNIVERSE}
\protect\label{Static}

In this case the expansion vanishes and consequently $\gamma (t)$ and
$\Delta (t)$ become constant. The above Hamiltonian reduces thus to:
\begin{equation}
{\cal H} = - \frac{1}{2 \mu^{2}} \hspace{0.1cm}P^{2} + \frac{1}{2}
\hspace{0.1cm}\omega^{2} \hspace{0.1cm} Q^{2}
\protect\label{h3}
\end{equation}
where $\mu$ and $\omega$ are obtained from eq.(\ref{h2}) for $\rho$ and
$A$ constant.

This is nothing but the oscillator Hamiltonian with an imaginary mass. We
recover then the well known result of the instability of Einstein\rq s
universe.

\section{FIERZ-LANCZOS POTENTIAL}
\protect\label{lanczos}

As it was remarked in a previous section, perturbations of conformally flat
spacetimes do not need\footnote{The reader should refer
to the above quoted gauge problem which has been widely discussed in the
literature (see the references given in the Introduction).} the complete
knowledge of all components of the
perturbed metric tensor $\delta g_{\mu\nu}$, although they certainly need to
take into account the Weyl conformal tensor, since all the observable
information we need is contained in it (namely, $\delta E_{ij}$ and
$\delta H_{ij}$).

Let us note at this point that the tensor $W_{\alpha\beta\mu\nu}$
can be expressed in terms of the 3-index Fierz-Lanczos potential tensor,
\cite{Fierz}, \cite{Lanczos}, that we will
denote by $L_{\alpha\beta\mu}$, and which deserves a careful analysis. Indeed,
one could consider $\delta L_{\alpha\beta\mu}$ as the good object for studying
linear perturbation theory, since as we shall see
it combines both $\delta\Sigma_{ij}$ and
$\delta a_{k}$ (which are alternative variables to describe $\delta E_{ij}$).

Before going into the perturbation-related details let us summarize here some
definitions and properties of $L_{\alpha\beta\mu}$, since the literature has
very few papers on this matter\footnote{This tensor was introduced in the
30\rq s to provide, in a similar way as the symmetric tensor
$\varphi_{\mu\nu}$
does --- in a more used approach --- an alternative description of spin-2
field in Minkowski background. In the 60\rq s Lanczos rediscovered it ---
without recognizing he was dealing with the same object --- as a Lagrange
multiplier in order to obtain the Bianchi identities in the context of
Einstein\rq s General Relativity. However
only recently \cite{NovelloNelson}, \cite{NovelloNelson2} a complete analysis
of Fierz-Lanczos object was undertaken
and it was discovered that its generic (Fierz) version describes not only one
but two spin-2 fields. The restriction to just a single spin-2 field is
usually called the Lanczos tensor. We will limit all our considerations here
to this restricted quantity.}.

\subsection{BASIC PROPERTIES}
\protect\label{basic}

In any 4-dimensional Riemannian geometry there exists a 3-index
tensor $L_{\alpha\beta\mu}$ which has the following symmetries:
\begin{equation}
L_{\alpha\beta\mu} + L_{\beta\alpha\mu} = 0
\protect\label{b1}
\end{equation}
\begin{equation}
L_{\alpha\beta\mu} + L_{\beta\mu\alpha} + L_{\mu\alpha\beta} = 0.
\protect\label{b2}
\end{equation}

With such $L_{\alpha\beta\mu}$ we may write the Weyl tensor in form of a
homogeneous expression in the potential expression, that is
\begin{eqnarray}
W_{\alpha\beta\mu\nu} & = & L_{\alpha\beta [\mu ;\nu ]} +
L_{\mu\nu[\alpha ;\beta ]} + \nonumber \\
& + & \frac{1}{2} [ L_{(\alpha\nu)} g_{\beta\mu} + L_{(\beta\mu)}
g_{\alpha\nu} - L_{(\alpha\mu)} g_{\beta\nu} -
L_{(\beta\nu)} g_{\alpha \mu}] + \nonumber \\
& + & \frac{2}{3}
{L^{\sigma\lambda}}_{  \sigma ;\lambda} \hspace{0.1cm}g_{\alpha\beta\mu\nu}
\protect\label{b3}
\end{eqnarray}
where
\begin{displaymath}
L_{\alpha\mu} \equiv {{L_{\alpha}}^{\sigma}}_{\mu ;\sigma} -
L_{\alpha ;\mu}
\end{displaymath}
and
\begin{displaymath}
L_{\alpha} \equiv {{L_{\alpha}}^{ \sigma}}_{\sigma}.
\end{displaymath}

Let us point out that, due to the above symmetry properties,
eqs.(\ref{b1}) and (\ref{b2}), Lanczos tensor has 20 degrees of freedom.
Since Weyl tensor has only 10 independent components, it follows that
there is a gauge symmetry involved. This gauge symmetry can be separated into
two classes:
\begin{equation}
\Delta^{(1)} L_{\alpha\beta\mu} = M_{\alpha} \hspace{0.1cm}g_{\beta\mu} -
M_{\beta} \hspace{0.1cm}g_{\alpha\mu}
\protect\label{b4}
\end{equation}
\begin{eqnarray}
\Delta^{(2)} L_{\alpha\beta\mu} & = & W_{\alpha\beta ;\mu} - \frac{1}{2}
W_{\mu\alpha ;\beta} + \frac{1}{2} W_{\mu\beta ;\alpha} \nonumber \\
& + & \frac{1}{2} g_{\mu\alpha} \hspace{0.1cm}
{{W_{\beta}}^{\lambda}}_{;\lambda} -
\frac{1}{2} g_{\mu\beta} \hspace{0.1cm}{{W_{\alpha}}^{\lambda}}_{;\lambda}
\protect\label{b5}
\end{eqnarray}
in which the vector $M_{\alpha}$ and the antisymmetric tensor
$W_{\alpha\beta}$ are arbitrary quantities.

\subsection{LANCZOS TENSOR FOR FRW GEOMETRY}
\protect\label{lanczos for frw}

The fact that Friedmann-Robertson-Walker geometry is conformally flat implies
that the associated Lanczos potential is nothing but a gauge. That is, we can
write the Lanczos potential for FRW geometry as
\begin{eqnarray}
L_{\alpha\beta\mu} & = & N_{\alpha} \hspace{0.1cm}g_{\beta\mu} -
N_{\beta} \hspace{0.1cm}g_{\alpha\mu}
+ F_{\alpha\beta ;\mu} - \frac{1}{2} F_{\mu\alpha ;\beta} \nonumber \\
& + & \frac{1}{2} F_{\mu\beta ;\alpha} + \frac{1}{2} g_{\mu\alpha}
\hspace{0.1cm}{{F_{\beta}}^{\lambda}}_{;\lambda} -
\frac{1}{2} g_{\mu\beta} \hspace{0.1cm}{{F_{\mu}}^{\lambda}}_{;\lambda}
\protect\label{b6}
\end{eqnarray}
for the arbitrary vector $N_{\alpha}$ and the antisymmetric tensor
$F_{\alpha\beta}$.

\subsection{PERTURBED FIERZ-LANCZOS TENSOR}
\protect\label{perturbed Fierz}

In the case we are examining in this paper (irrotational perturbations)
the perturbed Weyl tensor reduces to the form
\begin{equation}
\delta W_{\alpha\beta\mu\nu} = (\eta_{\alpha\beta\gamma\varepsilon}
\hspace{0.1cm} \eta_{\mu\nu\lambda\rho} - g_{\alpha\beta\gamma\varepsilon}
\hspace{0.1cm} g_{\mu\nu\lambda\rho}) V^{\gamma} V^{\lambda} \delta
E^{\varepsilon\rho}.
\protect\label{p1}
\end{equation}
since the magnetic part of Weyl tensor remains zero in this case.

It then follows that the perturbed electric tensor is given in terms of
Lanczos potential as:
\begin{eqnarray}
- \delta E_{ij} & = & \delta L_{0i[0;j]} + \delta L_{0j[0;i]} -
\frac{1}{2} \delta L_{(00)} \gamma_{ij} \nonumber \\
& - & \frac{1}{2} \delta L_{(ij)} +
\frac{2}{3} \delta {L^{\sigma \lambda}}_{\sigma ;\lambda} \gamma_{ij}.
\protect\label{p2}
\end{eqnarray}

Although the $L_{\alpha\beta\mu}$ tensor is not a unique well defined object
(since it has the gauge freedom we discussed above) we can use
some theorems (see \cite{NovelloVeloso}, \cite{LopezBonilla}) that enable one
to write $L_{\alpha\beta\mu}$ in terms of the associated
kinematic quantities of a given congruence of curves present in the
associated Riemannian manifold. Following these theorems and choosing
the case of irrotational perturbed matter it follows that
$\delta L_{\alpha\beta\mu}$ (the perturbed tensor of FRW background) is
given by
\begin{equation}
\delta L_{\alpha\beta\mu} = \delta \sigma_{\mu [\alpha} V_{\beta]} +
F(t) \delta a_{[\alpha} V_{\beta ]} V_{\mu}
\protect\label{p3}
\end{equation}
where
\begin{equation}
F(t) = 1 - \frac{1}{m} \frac{\Sigma}{\Psi} \left(\frac{2}{3} \Theta +
\frac{1}{2} \xi\right).
\protect\label{p4}
\end{equation}

In other words, the only non identically zero components of
$\delta L_{\alpha\beta\mu}$ are:
\begin{equation}
\delta L_{0k0} = - F(t) \Psi\hspace{0.1cm}Q_{k}
\protect\label{p5}
\end{equation}
and
\begin{equation}
\delta L_{0ij} = - \Sigma (t) \hat{Q}_{ij}
\protect\label{p6}
\end{equation}
that coincides with the previous results.

{}From what we have learned in the previous section, we can conclude that this
is not a univocal expression, that is, eqs.(\ref{p5}) and (\ref{p6}) are
obtained by a specific gauge choice.

Let us apply the above gauge transformation to the present
case. In the first gauge, eq.(\ref{b4}), we decompose vector $M_{\alpha}$ in
the spatial harmonics (scalar and vector):
\begin{equation}
M_{0} = M^{(1)}(t) \hspace{0.1cm}Q(x)
\protect\label{p7}
\end{equation}
\begin{equation}
M_{i} = M^{(2)}(t) \hspace{0.1cm}Q_{i}(x)
\protect\label{p8}
\end{equation}
and in the second gauge, eq.(\ref{b5}), we have
\begin{equation}
W_{0i} = W^{(1)}(t) \hspace{0.1cm}Q_{i}(x)
\protect\label{p9}
\end{equation}
and
\begin{equation}
W_{ij} = - \frac{1}{A^{2}} \hspace{0.1cm}\varepsilon_{ijk} \hspace{0.1cm}
W^{(2)}(t) \hspace{0.1cm}Q^{k}(x).
\protect\label{p10}
\end{equation}

To sum up, asking what is the Lanczos tensor for the perturbed FRW geometry is
one of those questions (like the one about the perturbed tensor
$\delta g_{\mu\nu}$) that should be avoided, since this quantity is
gauge-dependent. A good question to be asked should be --- as we remarked
before: What is the perturbation of Weyl tensor? This was precisely the
motivation of the previous section.

\section{CONCLUSION}
\protect\label{conclusion}

In this paper we have shown that the electric part of the Weyl tensor $E$ and
the shear $\Sigma$ can be taken as the two basic quantities which
describe the evolution of all perturbed quantities of FRW universe in the
irrotational case. For rotational perturbations the magnetic part of Weyl
tensor, $H$, and the vorticity $\Omega$ should be also included in the set
${\cal M}_{[A]}$. The proof of
this remark will be presented in a forthcoming paper.

We used the quasi-Maxwellian system of equations, which is equivalent to
Einstein\rq s equations, but is more convenient to treat perturbations in
conformally flat universes, e.g., FRW cosmology. We showed that it is possible
to reduce all the dynamics to a pair of equations in $\Sigma$ and $E$,
providing a dynamic planar system. A reparametrization of these variables
allows us then to establish a gauge-invariant Hamiltonian treatment for this
class of perturbation.

This suggests a natural way of quantization, in which $Q$ and $P$ become
operators of a Hilbert space. A simple look into the Hamiltonian
suggests that this quantization will give rise to single mode squeeze
states \cite{Grishchuk}. It is not difficult to see that introducting
the pair $(\Omega, H)$ will generate double mode squeeze states. This
analysis is now under development.

\vspace{2.0cm}

{\bf Acknowledgements}

M.C. Motta da Silva and R. Klippert would like to thank CNPq
for a grant. S.E. Jor\'{a}s would also like to thank CAPES for a grant.

\newpage

\section{APPENDIX --- QUASI-MAXWELLIAN EQUATIONS}
\protect\label{apb}

We list below the quasi-Maxwellian equations of gravity. They are obtained
from Bianchi identities as true dynamical equations which describe the
propagation of gravitational disturbances. Making use of Einstein\rq s
equations and the definition of Weyl tensor, Bianchi identities can be written
in an equivalent form as
\begin{eqnarray}
{W^{\alpha\beta\mu\nu}}_{;\nu} & = & \frac{1}{2}R^{\mu[\alpha ;\beta]} -
\frac{1}{12}g^{\mu[\alpha}R^{,\beta]} \nonumber \\
& = & - \frac{1}{2}T^{\mu[\alpha ;\beta]}
+ \frac{1}{6}g^{\mu[\alpha}T^{,\beta]}. \nonumber
\end{eqnarray}

Using the decomposition of Weyl tensor in terms of $E_{\alpha\beta}$ and
$H_{\alpha\beta}$ (see Section \ref{definitions}) and projecting apropriately,
Einstein\rq s equations can be written in a form which is similar to
Maxwell\rq s equations. There are 4 independent projections for the
divergence of Weyl tensor, namely:
\begin{displaymath}
\begin{array}{ll}
{W^{\alpha\beta\mu\nu}}_{;\nu} \hspace{0.1cm}V_{\beta} V_{\mu} \hspace{0.1cm}
{h_{\alpha}}^{\sigma} \\ \\
{W^{\alpha\beta\mu\nu}}_{;\nu} \hspace{0.1cm}
{\eta^{\sigma\lambda}}_{\alpha\beta} \hspace{0.1cm}V_{\mu}V_{\lambda} \\ \\
{W^{\alpha\beta\mu\nu}}_{;\nu} \hspace{0.1cm}{h_{\mu}}^{(\sigma} \hspace{0.1cm}
{\eta^{\tau)\lambda}}_{\alpha\beta} \hspace{0.1cm}V_{\lambda} \\ \\
{W^{\alpha\beta\mu\nu}}_{;\nu} \hspace{0.1cm}V_{\beta} \hspace{0.1cm}
h_{\mu (\tau}h_{\sigma )\alpha}.
\end{array}
\end{displaymath}

The unperturbed quasi-Maxwellian equations are thus given by:
\begin{eqnarray}
h^{\varepsilon\alpha} h^{\lambda\gamma} \hspace{0.1cm}
E_{\alpha\lambda;\gamma} & + & {\eta^{\varepsilon}}_{\beta\mu\nu} V^{\beta}
\hspace{0.1cm}H^{\nu\lambda} \hspace{0.1cm} {\sigma^{\mu}}_{\lambda} +
3H^{\varepsilon\nu} \hspace{0.1cm}\omega_{\nu} \nonumber \\
& = & \frac{1}{3} h^{\varepsilon\alpha} \hspace{0.1cm}\rho_{,\alpha}
+ \frac{\Theta}{3} q^{\varepsilon} - \frac{1}{2}
({\sigma^{\varepsilon}}_{\nu} - 3{\omega^{\varepsilon}}_{\nu})\hspace{0.1cm}
q^{\nu} \nonumber \\
& + & \frac{1}{2}\pi^{\varepsilon\mu} \hspace{0.1cm}a_{\mu} +
\frac{1}{2} h^{\varepsilon\alpha} \hspace{0.1cm}{{\pi_{\alpha}}^{\nu}}_{;\nu}
\protect\label{apb1}
\end{eqnarray}
\begin{eqnarray}
h^{\varepsilon\alpha} \hspace{0.1cm}h^{\lambda\gamma} \hspace{0.1cm}
H_{\alpha\lambda ;\gamma} & - & {\eta^{\varepsilon}}_{\beta\mu\nu}
\hspace{0.1cm}V^{\beta} \hspace{0.1cm}E^{\nu\lambda} \hspace{0.1cm}
{\sigma^{\mu}}_{\lambda} - 3E^{\varepsilon\nu} \hspace{0.1cm}\omega_{\nu}
\nonumber \\
& = & (\rho + p)\omega^{\varepsilon} - \frac{1}{2} \hspace{0.1cm}
\eta^{\varepsilon\alpha\beta\lambda}
\hspace{0.1cm}V_{\lambda} \hspace{0.1cm}q_{\alpha ;\beta} \nonumber \\
& + & \frac{1}{2} \hspace{0.1cm}\eta^{\varepsilon\alpha\beta\lambda}
(\sigma_{\mu\beta} +
\omega_{\mu\beta}) \hspace{0.1cm}{\pi^{\mu}}_{\alpha} \hspace{0.1cm}V_{\lambda}
\protect\label{apb2}
\end{eqnarray}
\begin{eqnarray}
{h_{\mu}}^{\varepsilon} {h_{\nu}}^{\lambda} \hspace{0.1cm}\dot{H}^{\mu\nu}
& + & \Theta H^{\varepsilon\lambda} - \frac{1}{2}{H_{\nu}}^{(\varepsilon}
{h^{\lambda )}_{\mu} \hspace{0.1cm}V^{\mu ;\nu} \nonumber \\
& + & \eta^{\lambda\nu\mu\gamma} \eta^{\varepsilon\beta\tau\alpha}
\hspace{0.1cm}V_{\mu} V_{\tau} \hspace{0.1cm}H_{\alpha\gamma} \hspace{0.1cm}
\Theta_{\nu\beta} \nonumber \\
& - & a_{\alpha} {E_{\beta}}^{(\lambda}
\eta^{\varepsilon )\gamma\alpha\beta} \hspace{0.1cm}V_{\gamma} \nonumber \\
& + & \frac{1}{2}{{E_{\beta}}^{\mu}}_{;\alpha} \hspace{0.1cm}
{h_{\mu}}^{(\varepsilon}\eta^{\lambda )\gamma\alpha\beta} \hspace{0.1cm}
V_{\gamma} \nonumber \\
& = & - \frac{3}{4}q^{(\varepsilon}\omega^{\lambda )} +
\frac{1}{2}h^{\varepsilon\lambda} \hspace{0.1cm}q^{\mu} \omega_{\mu}
\nonumber \\
& + & \frac{1}{4}{\sigma_{\beta}}^{(\varepsilon}
\eta^{\lambda )\alpha\beta\mu} \hspace{0.1cm}V_{\mu} q_{\alpha} \nonumber \\
& + & \frac{1}{4} h^{\nu(\varepsilon}\eta^{\lambda )\alpha\beta\mu}
\hspace{0.1cm}V_{\mu} \hspace{0.1cm}\pi_{\nu\alpha ;\beta}
\protect\label{apb3}
\end{eqnarray}
\begin{eqnarray}
{h_{\mu}}^{\varepsilon} {h_{\nu}}^{\lambda} \hspace{0.1cm}\dot{E}^{\mu\nu}
& + & \Theta E^{\varepsilon\lambda} - \frac{1}{2} {E_{\nu}}^{(\varepsilon}
{h^{\lambda )}}_{\mu} \hspace{0.1cm}V^{\mu ;\nu} \nonumber \\
& + & \eta^{\lambda\nu\mu\gamma} \eta^{\varepsilon\beta\tau\alpha}
\hspace{0.1cm}V_{\mu}V_{\tau} \hspace{0.1cm}E_{\alpha\gamma} \Theta_{\nu\beta}
+ a_{\alpha} {H_{\beta}}^{(\lambda}
\eta^{\varepsilon )\gamma\alpha\beta} \hspace{0.1cm}V_{\gamma} \nonumber \\
& - & \frac{1}{2} {{H_{\beta}}^{\mu}}_{;\alpha} \hspace{0.1cm}
{h_{\mu}}^{(\varepsilon}\eta^{\lambda )\gamma\alpha\beta} \hspace{0.1cm}
V_{\gamma} \nonumber \\
& = & \frac{1}{6}h^{\varepsilon\lambda} ({q^{\mu}}_{;\mu} -
q^{\mu} a_{\mu} - \pi^{\nu\mu} \sigma_{\mu\nu}) \nonumber \\
& - & \frac{1}{2}(\rho + p) \sigma^{\varepsilon\lambda} + \frac{1}{2}
q^{(\varepsilon} a^{\lambda )} \nonumber \\
& - & \frac{1}{4} h^{\mu (\varepsilon} h^{\lambda )\alpha} \hspace{0.1cm}
q_{\mu ;\alpha} + \frac{1}{2}{h_{\alpha}}^{\varepsilon}{h_{\mu}}^{\lambda}
\hspace{0.1cm}\dot{\pi}^{\alpha\mu} \nonumber \\
& + & \frac{1}{4}{\pi_{\beta}}^{(\varepsilon} \sigma^{\lambda )\beta} -
\frac{1}{4}{\pi_{\beta}}^{(\varepsilon} \omega^{\lambda )\beta} + \frac{1}{6}
\Theta\pi^{\varepsilon\lambda}.
\protect\label{apb4}
\end{eqnarray}

The contracted Bianchi identities and Einstein\rq s equations give the
conservation law
\begin{displaymath}
{T^{\mu\nu}}_{;\nu} = 0.
\end{displaymath}
Projecting it both in the parallel and the orthogonal subspaces we obtain:
\begin{displaymath}
\begin{array}{ll}
{T^{\mu\nu}}_{;\nu} V_{\mu} = 0 \\ \\
{T^{\mu\nu}}_{;\nu} {h_{\mu}}^{\alpha} = 0
\end{array}
\end{displaymath}
which give the following equations:
\begin{equation}
\dot{\rho} + (\rho + p)\Theta + \dot{q}^{\mu} V_{\mu} + {q^{\alpha}}_{;\alpha}
- \pi^{\mu\nu}\Theta_{\mu\nu} = 0
\protect\label{apb5}
\end{equation}
\begin{eqnarray}
(\rho + p)a_{\alpha} & - & p_{,\mu}{h^{\mu}}_{\alpha} +
\dot{q}_{\mu} {h^{\mu}}_{\alpha} + \Theta q_{\alpha} \nonumber \\
& + & q^{\nu}\Theta_{\alpha\nu} + q^{\nu}\omega_{\alpha\nu} +
{{\pi_{\alpha}}^{\nu}}_{;\nu} + \pi^{\mu\nu}\Theta_{\mu\nu}V_{\alpha} = 0
\protect\label{apb6}
\end{eqnarray}
and, from the definition of Riemann curvature tensor
\begin{displaymath}
V_{\mu ;\alpha ;\beta} - V_{\mu ;\beta ;\alpha} =
R_{\mu\varepsilon\alpha\beta} V^{\varepsilon}
\end{displaymath}
we obtain the equations of motion for the unperturbed kinematical quantities
as:
\begin{equation}
\dot{\Theta} + \frac{\Theta^{2}}{3} + 2\sigma^{2} + 2\omega^{2} -
{a^{\alpha}}_{;\alpha} = R_{\mu\nu} V^{\mu} V^{\nu}
\protect\label{apb7}
\end{equation}
\begin{eqnarray}
{h_{\alpha}}^{\mu} {h_{\beta}}^{\nu} \hspace{0.1cm}\dot{\sigma}_{\mu\nu}
& + & \frac{1}{3}h_{\alpha\beta} (-2\omega^{2} - 2\sigma^{2} +
{a^{\lambda}}_{;\lambda}) + a_{\alpha} a_{\beta} \nonumber \\
& - & \frac{1}{2}{h_{\alpha}}^{\mu}{h_{\beta}}^{\nu} \hspace{0.1cm}
(a_{\mu ;\nu} + a_{\nu ;\mu}) + \frac{2}{3}\Theta\sigma_{\alpha\beta} +
\sigma_{\alpha\mu}{\sigma^{\mu}}_{\beta} +
\omega_{\alpha\mu}{\omega^{\mu}}_{\beta}
\nonumber \\
& = & R_{\alpha\varepsilon\beta\nu}V^{\varepsilon}V^{\nu} - \frac{1}{3}
R_{\mu\nu}V^{\mu}V^{\nu}h_{\alpha\beta}
\protect\label{apb8}
\end{eqnarray}
\begin{eqnarray}
{h_{\alpha}}^{\mu} {h_{\beta}}^{\nu} \hspace{0.1cm}\dot{\omega}_{\mu\nu}
& - & \frac{1}{2}{h_{\alpha}}^{\mu}{h_{\beta}}^{\nu}(a_{\mu ;\nu} -
a_{\nu ;\mu}) + \frac{2}{3}\Theta\omega_{\alpha\beta} \nonumber \\
& + & \sigma_{\alpha\mu}{\omega^{\mu}}_{\beta} - \sigma_{\beta\mu}
{\omega^{\mu}}_{\alpha} = 0.
\protect\label{apb9}
\end{eqnarray}

We also obtain from the definition of $R_{\alpha\beta\mu\nu}$ three
constraint equations:
\begin{equation}
\frac{2}{3}\Theta_{,\mu}{h^{\mu}}_{\lambda} - ({\sigma^{\alpha}}_{\gamma} +
{\omega^{\alpha}}_{\gamma})_{;\alpha}{h^{\gamma}}_{\lambda} - a^{\nu}
(\sigma_{\lambda\nu} + \omega_{\lambda\nu}) = R_{\mu\nu}V^{\mu}
{h^{\nu}}_{\lambda}
\protect\label{apb10}
\end{equation}
\begin{equation}
{\omega^{\alpha}}_{;\alpha} + 2\omega^{\alpha} \hspace{0.1cm}a_{\alpha} = 0
\protect\label{apb11}
\end{equation}
\begin{equation}
- \frac{1}{2} \hspace{0.1cm}{h_{\tau}}^{\varepsilon} \hspace{0.1cm}
{h_{\lambda}}^{\alpha} \hspace{0.1cm}{\eta_{\varepsilon}}^{\beta\gamma\nu}
\hspace{0.1cm}V_{\nu} \hspace{0.1cm}(\sigma_{\alpha\beta} +
\omega_{\alpha\beta})_{;\gamma} + a_{(\tau} \hspace{0.1cm}\omega_{\lambda )}
= H_{\tau\lambda}.
\protect\label{apb12}
\end{equation}

These results constitute a set of 12 equations which will be used to describe
the evolution of small perturbations in FRW background. Writing all the
perturbed quantities in the form
\begin{displaymath}
X_{(perturbed)} = X_{(background)} + \delta X
\end{displaymath}
and after straightforward manipulations we finally obtain the perturbed
equations from the set of equations (\ref{apb1})-(\ref{apb12}) as:
\begin{eqnarray}
(\delta E^{\mu\nu})^{\bullet} \hspace{0.1cm}
{h_{\mu}}^{\alpha}{h_{\nu}}^{\beta} & + &
\Theta \hspace{0.1cm}(\delta E^{\alpha\beta}) - \frac{1}{2} (\delta
{E_{\nu}}^{(\alpha}){h^{\beta )}}_{\mu}
\hspace{0.1cm}V^{\mu ;\nu} \nonumber \\
& + & \frac{\Theta}{3}\eta^{\beta\nu\mu\varepsilon} \hspace{0.1cm}
\eta^{\alpha\gamma\tau\lambda} \hspace{0.1cm}V_{\mu}V_{\tau}
(\delta E_{\varepsilon\lambda}) \hspace{0.1cm}h_{\gamma\nu} \nonumber \\
& - & \frac{1}{2} (\delta {{H_{\lambda}}^{\mu}})_{;\gamma} \hspace{0.1cm}
{h_{\mu}}^{(\alpha}\eta^{\beta )\tau\gamma\lambda}V_{\tau} \nonumber \\
& = & - \frac{1}{2}(\rho + p)\hspace{0.1cm}(\delta \sigma^{\alpha\beta})
\nonumber \\
& + & \frac{1}{6} \hspace{0.1cm}h^{\alpha\beta} \hspace{0.1cm}(\delta
q^{\mu})_{;\mu} - \frac{1}{4} \hspace{0.1cm}h^{\mu (\alpha} h^{\beta )\nu}
\hspace{0.1cm}(\delta q_{\mu})_{;\nu} \nonumber \\
& + & \frac{1}{2} \hspace{0.1cm}h^{\mu(\alpha} h^{\beta )\nu} \hspace{0.1cm}
(\delta\Pi_{\mu\nu})^{\bullet} + \frac{1}{6} \hspace{0.1cm}\Theta
\hspace{0.1cm}(\delta\Pi^{\alpha\beta})
\protect\label{apb13}
\end{eqnarray}
\begin{eqnarray}
({\delta H^{\mu\nu})^{\bullet} \hspace{0.1cm}
{h_{\mu}}^{\alpha}{h_{\nu}}^{\beta} & + & \Theta \hspace{0.1cm}
(\delta H^{\alpha\beta}) - \frac{1}{2} (\delta {H_{\nu}}^{(\alpha})
{h^{\beta )}_{\mu} \hspace{0.1cm}V^{\mu ;\nu} \nonumber \\
& + & \frac{\Theta}{3} \eta^{\beta\nu\mu\varepsilon} \hspace{0.1cm}
\eta^{\alpha\lambda\tau\gamma} \hspace{0.1cm}V_{\mu}V_{\tau} \hspace{0.1cm}
(\delta H_{\varepsilon\gamma}) \hspace{0.1cm}h_{\lambda\nu} \nonumber \\
& - & \frac{1}{2} (\delta {{E_{\lambda}}^{\mu}})_{;\tau} \hspace{0.1cm}
{h_{\mu}}^{(\alpha}\eta^{\beta )\tau\gamma\lambda} \hspace{0.1cm}V_{\gamma}
\nonumber \\
& = & \frac{1}{4} \hspace{0.1cm}h^{\nu (\alpha}
\eta^{\beta )\varepsilon\tau\mu} \hspace{0.1cm}V_{\mu} \hspace{0.1cm}
(\delta\Pi_{\nu\varepsilon})_{;\tau}
\protect\label{apb14}
\end{eqnarray}
\begin{equation}
(\delta H_{\alpha\mu})_{;\nu} h^{\alpha\varepsilon} h^{\mu\nu} = (\rho + p)
\hspace{0.1cm}(\delta\omega^{\varepsilon}) - \frac{1}{2} \hspace{0.1cm}
\eta^{\varepsilon\alpha\beta\mu}
\hspace{0.1cm}V_{\mu} \hspace{0.1cm}(\delta q_{\alpha})_{;\beta}
\protect\label{apb15}
\end{equation}
\begin{eqnarray}
(\delta E_{\alpha\mu})_{;\nu} h^{\alpha\varepsilon}
\hspace{0.1cm}h^{\mu\nu} & = &
\frac{1}{3} (\delta\rho)_{,\alpha} h^{\alpha\varepsilon} - \frac{1}{3}
\dot{\rho} \hspace{0.1cm}(\delta V^{\varepsilon}) \nonumber \\
& - & \frac{1}{3} \hspace{0.1cm}\rho_{,0} \hspace{0.1cm}(\delta V^{0})
\hspace{0.1cm}V^{\varepsilon} \nonumber \\
& + & \frac{1}{2} \hspace{0.1cm}{h^{\varepsilon}}_{\alpha} \hspace{0.1cm}
(\delta\Pi^{\alpha\mu})_{;\mu} + \frac{\Theta}{3} \hspace{0.1cm}(\delta
q^{\varepsilon})
\protect\label{apb16}
\end{eqnarray}
\begin{equation}
(\delta\Theta)^{\bullet} + \frac{2}{3}\Theta \hspace{0.1cm}(\delta\Theta) -
(\delta {a^{\alpha}})_{;\alpha} = - \frac{(1 + 3\lambda)}{2} \hspace{0.1cm}
(\delta\rho)
\protect\label{apb17}
\end{equation}
\begin{eqnarray}
(\delta\sigma_{\mu\nu})^{\bullet} & + & \frac{1}{3}h_{\mu\nu}
(\delta {a^{\alpha}})_{;\alpha} - \frac{1}{2}(\delta a_{(\alpha})_{;\beta)}
\hspace{0.1cm}{h_{\mu}}^{\alpha} \hspace{0.1cm}{h_{\nu}^{\beta}
\nonumber \\
& + & \frac{2}{3}\Theta \hspace{0.1cm}(\delta\sigma_{\mu\nu}) = - (\delta
E_{\mu\nu}) - \frac{1}{2} (\delta\Pi_{\mu\nu})
\protect\label{apb18}
\end{eqnarray}
\begin{equation}
(\delta\omega^{\mu})^{\bullet} + \frac{2}{3}\Theta \hspace{0.1cm}
(\delta\omega^{\mu})
= \frac{1}{2} \eta^{\alpha\mu\beta\gamma} \hspace{0.1cm}(\delta
a_{\beta})_{ ;\gamma} \hspace{0.1cm}V_{\alpha}
\protect\label{apb19}
\end{equation}
\begin{eqnarray}
\frac{2}{3}(\delta\Theta)_{,\lambda} \hspace{0.1cm}{h^{\lambda}}_{\mu}
& - & \frac{2}{3}\dot{\Theta} \hspace{0.1cm} (\delta V_{\mu}) + \frac{2}{3}
\hspace{0.1cm}\dot{\Theta} \hspace{0.1cm}(\delta V^{0}) \hspace{0.1cm}
{\delta_{\mu}}^{0} \nonumber \\
& - & {(\delta {\sigma^{\alpha}}_{\beta} +
\delta {\omega^{\alpha}}_{\beta})}_{;\alpha} {h^{\beta}}_{\mu} = - (\delta
q_{\mu})
\protect\label{apb20}
\end{eqnarray}
\begin{equation}
(\delta {\omega^{\alpha}})_{;\alpha} = 0
\protect\label{apb21}
\end{equation}
\begin{equation}
(\delta H_{\mu\nu}) = - \frac{1}{2} \hspace{0.1cm}
{h^{\alpha}}_{(\mu} \hspace{0.1cm}
{h^{\beta}}_{\nu )} ((\delta\sigma_{\alpha\gamma})_{;\lambda} +
(\delta\omega_{\alpha\gamma})_{;\lambda}) \hspace{0.1cm}
{\eta_{\beta}}^{\varepsilon\gamma\lambda} \hspace{0.1cm}V_{\varepsilon}
\protect\label{apb22}
\end{equation}
\begin{equation}
(\delta\rho)^{\bullet} + \Theta \hspace{0.1cm}(\delta\rho + \delta p)
+ (\rho + p) \hspace{0.1cm}(\delta\Theta) + (\delta q^{\alpha})_{;\alpha} = 0
\protect\label{apb23}
\end{equation}
\begin{eqnarray}
\dot{p} \hspace{0.1cm}(\delta V_{\mu}) & + & p_{,0} \hspace{0.1cm}
(\delta V^{0}) \hspace{0.1cm}{\delta_{\mu}}^{0} - (\delta p)_{,\beta}
\hspace{0.1cm}{h^{\beta}}_{\mu} + (\rho + p) \hspace{0.1cm}(\delta a_{\mu})
\nonumber \\
& + & h_{\mu\alpha} (\delta q^{\alpha})^{\bullet} +  \frac{4}{3} \Theta
\hspace{0.1cm}(\delta q_{\mu}) + h_{\mu\alpha}
\hspace{0.1cm}(\delta \pi^{\alpha\beta})_{;\beta} = 0.
\protect\label{apb24}
\end{eqnarray}


\newpage

\begin{thebibliography}{99}
 \bibitem{Lifshitz}E. M. Lifshitz and I.M. Khalatnikov, Adv.Phys. {\bf 12},
(1963), 185.
 \bibitem{Bardeen}J. Bardeen, Phys. Rev. D {\bf 22}, 8 (1980), 1882.
 \bibitem{Jones}B. J. T. Jones, Rev. Mod. Phys. {\bf 48}, 1 (1976), 107.
 \bibitem{Hawking}S. W. Hawking, Ap. J. {\bf 145} (1966), 544.
 \bibitem{Olson}D. W. Olson, Phys. Rev. D {\bf 14}, (1976), 327.
 \bibitem{Vishniac}J. C. Hwang and E.T. Vishniac, Ap. J. {\bf 353}, (1990),
1-20.
 \bibitem{BrandenbergerKhan}R. Brandenberger, R. Khan and W. H. Press, Phys.
Rev. D {\bf 28}, (1983), 1809.
 \bibitem{MukhanovBrand}V. F. Mukhanov, H. A. Feldmann and R. H. Brandenberger,
 Phys. Rep. {\bf 215}, (1992).
 \bibitem{Jordan et al}P. Jordan, J. Ehlers and R. Sachs, Akad. Wiss. Lt.
Meinz Abh. Math. Nat. Kl. {\bf 1}, (1961), 3.
 \bibitem{Novello e Salim}M. Novello, and J. M. Salim, Fund. of Cosmic Phys.
{\bf 8}, (1983), 201-342.
 \bibitem{Stewart}J. M. Stewart and M. Walker, Proc. R. Soc. London, {\bfA341},
(1974), 49-74.
 \bibitem{Debever}R. Debever, Cahiers de Physique {\bf 168}, (1964).
 \bibitem{Israel}W. Israel, An. Phys. {\bf 100}, (1976), 310-331.
 \bibitem{NovelloSalim}M. Novello, J. M. Salim and H. Heintzmann {\it in J.
Leite Lopes Festschrift}, 1988, Ed. S. Joffily.
 \bibitem{Ellis et al}G. F. R. Ellis and M. Bruni, Phys.Rev. D {\bf 40}, 6
(1989), 1804.
 \bibitem{Grishchuk}L. P. Grishchuk and Yu V. Sidorov, Phys. Rev. D, {\bf 42},
 (1990), 3413.
 \bibitem{Fierz}M. Fierz, Helv. Phys. Acta {\bf 12}, (1939), 379.
 \bibitem{Lanczos}C. Lanczos, Rev. Mod. Phys. {\bf 34}, (1962), 379.
 \bibitem{NovelloNelson}M. Novello and N. Pinto-Neto, Fortschrift der
Physik {\bf 40}, (1992), 173.
 \bibitem{NovelloNelson2}M. Novello and N. Pinto-Neto, Fortschrift der
Physik {\bf 40}, (1992), 195.
 \bibitem{NovelloVeloso}M. Novello and A. L. Velloso, Gen. Rel. Grav. {\bf 19},
(1987), 1251.
 \bibitem{LopezBonilla}G. Ares de Parga, O. Chavoya, J.L. L\'{o}pez Bonilla,
J. Math. Phys. {\bf 30}, (1989), 1294.
\end{thebibliography}
\end{document}